# Identifying Optimal Market Choices to Increase the Profitability of Coffee Farmers in Sultan Kudarat through Modeling and Scenario Analysis

**Novy Aila B. Rivas[1][2]\*, Giovanna Fae R. Oguis[1][2], Alex John C. Labanon[1][2],**

**El Veena Grace A. Rosero[1][2], Jon Henly O. Santillan[1][2], and Larry N. Digal[1][3]**

[1]Agri-Aqua Value Chain Laboratory

[2]Department of Mathematics, Physics, and Computer Science

[3]School of Management

University of the Philippines Mindanao, Davao City, 8000 Philippines

\*Corresponding Author: groguis@up.edu.ph; (082) 293 0303

## ABSTRACT

Coffee farmers face a range of challenges that reduce their income. These include low productivity, lack of capital and limited access to credit, knowledge of farming technologies, high production costs, and inability to meet market requirements in terms of quality and volume. A smaller farm size also affects the earnings of coffee farmers, i.e., as coffee farm sizes decrease, the smallholder coffee farmers earn less. This study focuses on the coffee chain of Sultan Kudarat – the top producer of coffee in the Philippines, where most of the farmers are smallholders. Coffee farmers in this area allocate their harvested cherries as fresh cherries, dried cherries, and green coffee beans to five market outlets: Nestle Philippines, local traders, growers' association, direct selling, and other markets not mentioned (e.g., coffee shops and hotels). Choosing the best market to sell their product is a problem for farmers, especially when there are restrictions in marketing their coffee. Hence, a supply chain network design (SCND) model and simulation are developed to investigate the changes in the profits of coffee farmers as they market their products, whether to be sold as fresh cherries, dried cherries, or processed into green coffee beans before marketing to the market outlets mentioned above, based on the average annual costs affecting the production, primary processing, and market prices of coffee products. Assuming that the annual coffee yield per tree and the average prices of coffee products in different markets are constant, the simulations show that farmers can gain a maximum annual profit of Php 47,590.00 per hectare if they sell all fresh cherries to a market with an average selling price of around Php 32.00 per kilogram. Farmers can also gain a positive annual profit averaging Php 43,883.85 per hectare if they sell all dried cherries. However, results show that if farmers decide to produce and sell all green coffee beans, the farmers gain a negative profit due to the additional annual average dehulling cost of Php 13.63 per tree and the minimal difference in average selling prices between dried cherries and green coffee beans in different markets. Furthermore, members of the grower's association producing dried cherries can gain a maximum annual profit of Php 46,550.71 per hectare by allocating the required 70% of their product to the association and 30% to the other markets. In contrast, selling 30% of green coffee beans to any market generates a negative profit. Based on the data and assumptions used in the simulation and given the farmer's current resources and farm practices, selling fresh or dried cherries is recommended to gain positive profit. Finally, this approach was inspired by the developed SCND model for the Cavendish banana commodity in the Davao Region, which serves as a diagnostic tool to assess the profitability of a commodity among different market options. Therefore, the developed model can be modified and used for regular coffee farms and other commodities.

*Corresponding Author: groguis@up.edu.ph; (082) 293 0303

**INTRODUCTION**

The Philippines, situated within the 'bean belt' or 'coffee belt,' is blessed with a climate and soil conditions conducive to coffee cultivation (Philippine Coffee Board, Inc., 2021). Coffee is one of the country's high-value crops and is considered the second-largest agricultural export, which generates an annual revenue of over $100 million (Philippine Coffee Board, nd). Coffee belongs to the genus Coffea- a member of the *Rubiaceae* family that holds more than 90 different species (Murthy & Naidum, 2012). However, in the Philippines, only four varieties are commercially viable (Philippine Coffee Industry

*Corresponding Author: groguis@up.edu.ph; (082) 293 0303

Roadmap, 2022). Robusta variety accounts for 76.5% of the total coffee production in the country, followed by arabica, excelsa, and liberica with 16.7%, 5.7%, and 1.1%, respectively (PSA, nd).

Mindanao is the top coffee-producing island in the country producing around 83.63% of dried coffee cherries per year (*Philippine Coffee Industry Roadmap 2021-2025 - Philippine Council for Agriculture and Fisheries*, 2022). It is also where four of the top 5 producing regions are located, SOCCSKSARGEN with a total dried cherries production of 35.5%, followed by Davao Region (17.85%), BARMM (17.42%), and Northern Mindanao (9.23%), respectively (PSA, 2020). Moreover, 95% of coffee farms in the Philippines measuring less than 5 hectares are owned by smallholder farmers. Specifically in Mindanao, smallholder farmers are vital in coffee production since they are the main producers of coffee on the island (DA & PCI, 2022; Philippine Coffee Industry Roadmap 2017-2022).

Over the past five years, coffee production has declined by 16.17% tons of dried cherries from 2015 to 2020 (PSA, nd). The downward trend of coffee production correlates to the declining number of bearing coffee trees, from 77.73M bearing trees in 2016 to 74.50M bearing trees in 2019. Moreover, the yield per bearing tree also decreased by 2.09% per year over the last 10 years. The low productivity is directly related to the aging coffee trees, limited rejuvenations, and poor farm management (PSA, 2021; PCI, 2020). Coffee farmers are reported to gain low income or worse even negative profit due to the challenges they face in coffee production. This includes the high cost of farm inputs and labor, lack of capital and limited access to credit, limited knowledge of appropriate coffee farming technologies, the effect of climate change, and natural calamities which will drive coffee farmers to cut trees and crop shifting (PCI, 2017).

In the year 2020, the Philippine Coffee Industry reported that there was an increase in the area planted with coffee trees by 1.1%. This is due to the increase of coffee growers and new key players in coffee chains, such as processing and manufacturing of specialty coffee, the increasing number of coffee shops, and rising household consumers of specialty coffee (PCI, 2020). This indicates that there is also an increase in market choices for coffee farmers.

Coffee is one of the most widely consumed beverages in the world and one of the most traded commodities globally (FAO; Haile & Kang, 2020). Around 30%-40% of the world's population consumes coffee every day which is about 2 billion cups (British Coffee Association, nd; Coffee Stats, 2022). In the Philippines, nine out of ten households have coffee in their pantries and eight out of ten adults drink an average of 2.5 cups of coffee every day (Lelis et al, 2023).

In 2017, the number of coffee shops in the Philippines has been growing steadily by a 3% increase in outlets. According to the study about the performance of the country's coffee manufacturing firm perspective of Consignado & Dimaculangan (2023), the threat of potential entrants in the coffee industry is high indicating that new companies can quickly enter the market. However, despite the increase of new key players in the coffee chain, farmers are still unable to capture sufficient value to commensurate with their hard labor in coffee farming (Habaradas et al, 2021).

*Corresponding Author: groguis@up.edu.ph; (082) 293 0303

The Value chain of the coffee industry involves five stages before reaching the cups of coffee lovers: production, primary processing, trading/consolidation, secondary processing, and marketing (Worldbank, 2018). However, smallholder farmers typically conclude the chain at the primary processing stage before trading it to the market, primarily due to the limited knowledge and a lack of facilities for secondary processing. Coffee producers usually market their coffee products, such as fresh cherries (FC) or green coffee beans (GCB), to traders, small/large processors, large companies, and specialty coffee shops. Despite having various market choices, smallholder farmers still encounter difficulties selling their products at high prices due to limited access to industrial and consumer markets. As a result, farmers are forced to sell their produce to middlemen often at very low prices, partly due to the quality of beans they produce, causing low income in selling coffee beans.

Identifying optimal market choices is crucial to increase the profitability of coffee farmers. One way to understand the coffee chain is by modeling its network using the supply chain network design (SCND) technique.

SCND modeling aims to optimize operational resources by accounting for various opportunities and constraints along the chain (Alzaman et al, 2018). This method is used in the optimization of resource allocation in different types of production processes (Oguis et al, 2022). The resources and operations/practices of smallholder farmers in Sultan Kudarat, the largest producer of coffee in the Philippines, serve as the basis for constructing the SCND model of coffee farmers. Scenario analysis is then performed using Python programming language to aid farmers in their decision-making process, providing them with optimal market choices for marketing their coffee beans and ultimately increasing their profits.

*Corresponding Author: groguis@up.edu.ph; (082) 293 0303

MATERIALS AND METHODS

### Dataset

The dataset used in the model is gathered by the Mahintana Foundation Inc., a non-governmental organization based in Sultan Kudarat, and obtained through the Inclusive Value Chain Project in UP Mindanao. The coffee farms in the data are situated in the municipalities of Kalamansig and Lebak, Sultan Kudarat. The raw dataset has a total of 2,925 surveyed coffee farmers in Sultan Kudarat, the top producer of coffee in the Philippines, where 2,886 cultivated robusta coffee variety, 37 arabica, and 2 farmers planted excelsa variety. The majority of coffee farmers in the area are small-scale with an average land area planted with coffee of 1.90 hectares. However, only 2,112 data points are used in the analysis, mainly respondents that cultivate robusta and are using strip picking methods during harvest.

### Supply chain network design (SCND)

Supply chain network design represents an integrated configuration of supply, manufacturing, and demand sub-systems, addressing the strategic decisions of each segment of the chain (Baghalian, Rezapour, & Farahani, 2013). This modeling approach is designed to optimize operation resources, focusing on strategic decision-making (Alzaman, Zhang, & Diabat, 2018).

*Corresponding Author: groguis@up.edu.ph; (082) 293 0303

Figure 1 illustrates the flow of the methodology for developing the coffee SCND model. The initial step involves conducting a literature review on the supply chain and value chain of coffee. This aims to identify key actors, factors, and parameters to be used in the model. Subsequently, a comprehensive framework detailing the production of coffee, from tree to market (Figure 2) was created. This framework depicts the various stages involved in processing coffee beans until they reach the marketing stage. The chain presented in Figure 2 can be modified into a specific chain depending on the specific practices of coffee farmers in a certain area. For instance, Figure 3 illustrates the coffee chain in Sultan Kudarat.

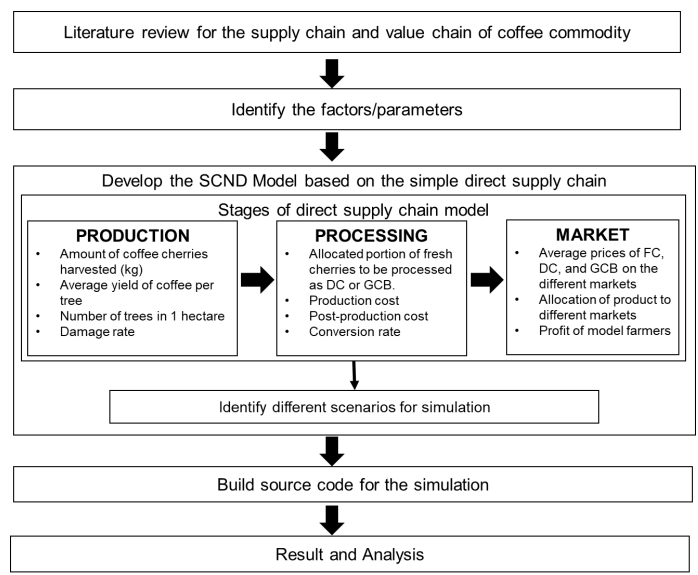

Figure 1. Flow of the methodology of coffee SCND model.

*Corresponding Author: groguis@up.edu.ph; (082) 293 0303

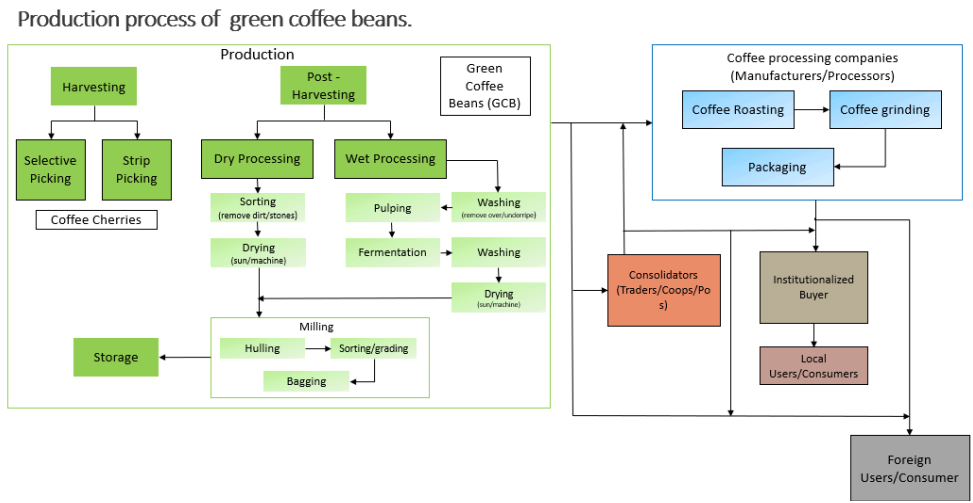

Figure 2. The production process of coffee from farm to market.

The methodology employed in developing the SCND model is adapted from a study exploring the profitability of small-scale Cavendish banana farmers through a model simulation approach by Mata et al (2020). After identifying the key actors in the coffee chain, a general model was formulated based on the direct supply chain model of the coffee chain. The direct supply chain of coffee is composed of three stages, production, processing, and market (Figure 1).

In the production stage, data on the amount of coffee cherries harvested in kilograms, the average yield of coffee per tree, the number of trees per hectare, and the damage rate incurred during harvesting were used. In the processing stage, calculations were made for the allocated portion of fresh cherries to be processed as dried cherries or green coffee beans, and production and post-production costs were determined using the available secondary data provided. Finally, the farmers' profit depends upon the allocation and average prices of coffee products to the identified market outlets.

*Corresponding Author: groguis@up.edu.ph; (082) 293 0303

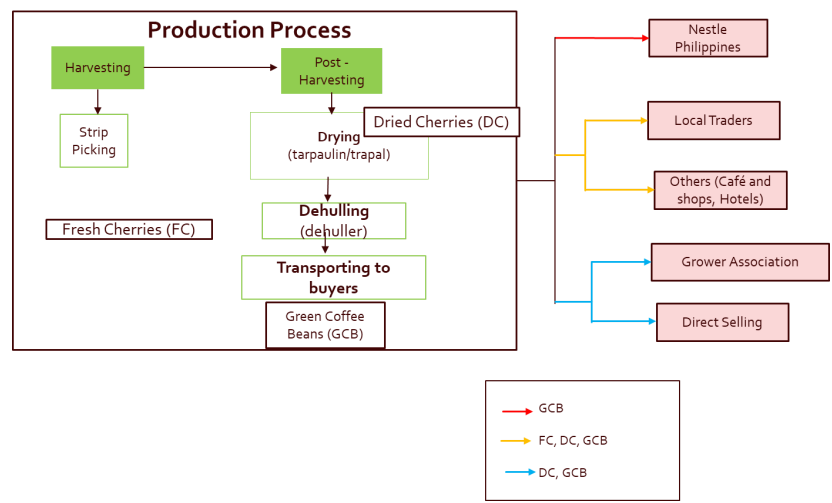

Figure 3. Coffee production of coffee farmers in the Kalamansig and Lebak, Sultan Kudarat municipalities.

***Coffee Chain in Sultan Kudarat***

The production process of coffee farmers based on the dataset is presented in Figure 3. The majority of the farmers used the strip picking method in harvesting their coffee cherries and used the natural processing method in processing their coffee cherries into green coffee beans. Natural processing is the traditional method of processing coffee cherries to green coffee beans. This process involves directly drying the harvested cherries under the sun or using a machine dryer. However, in Sultan Kudarat, the majority of smallholder farmers used trapal/tarpaulin to dry their beans under the sun. After the dried cherries reach their ideal moisture content, they will then undergo a dehulling process where the husk of the dried cherries is removed to obtain the green coffee beans.

Based on the dataset, there are 5 identified markets in the site, Nestle Philippines, local traders, grower association/cooperative, direct selling, and other markets such as café, coffee shops, and hotels. Additionally, coffee farmers sell 3 product types: fresh cherries

*Corresponding Author: groguis@up.edu.ph; (082) 293 0303

(FC), dried cherries (DC), and green coffee beans (GCB). Each product type has corresponding costs incurred. For instance, the cost of producing fresh cherries are the input cost (fertilizer), labor costs (fertilizer application, weeding, and pruning), Harvesting costs, transportation cost, and good agricultural practices (GAP) cost. If the farmer sells dried cherry, then there is another cost incurred for drying, and additional dehulling cost for processing and producing GCB. The parameter values used in the analysis for the costs are presented in Table 1.

Table 1. Coffee production activity/practices cost in 1 ha based on the Sultan Kudarat dataset

| Product Type | | | Costs incurred | Average cost (Php/tree) | SD |
|---|---|---|---|---|---|
| GCB | DC | FC | Fertilizer | 0.86 | 5.1848 |
| | | | Fertilizer application | 1.20 | 5.6521 |
| | | | Pruning | 2.38 | 4.2322 |
| | | | Weeding | 3.38 | 5.7509 |
| | | | Harvesting | 5.98 | 10.4203 |
| | | | Transportation | 2.35 | 3.4235 |
| | | | GAP | 15.20 | 21.1903 |
| | | | Drying | 0.56 | 1.8221 |
| | | | Dehulling | 2.51 | 2.9458 |

Table 2. Average prices of coffee products sold to different market outlets.

| Market Outlets | Product Type | | | | | |
|---|---|---|---|---|---|---|
| | Average Price (Php/Kg) | | | Standard Dev | | |
| | FC | DC | GCB | FC | DC | GCB |
| Nestle | 0 | 0 | 75.41 | 0 | 0 | 3.05956 |
| Local Traders | 32.5 | 70.7 | 69.7 | 27.157 | 4.51731 | 9.54011 |
| Grower Association | 0 | 75 | 74.5 | 0 | 0 | 0 |
| Direct Selling | 0 | 72.96 | 70.24 | 0 | 3.05956 | 3.65911 |
| Other Markets | 12 | 75 | 70.12 | 0 | 0 | 2.25832 |

* Value 0 in the prices indicates that the market does not buy that type of product.

*Corresponding Author: groguis@up.edu.ph; (082) 293 0303

Table 3. Parameters used in the model and its values.

| Parameter | Values | Description | Source |
|---|---|---|---|
| $\Lambda_t$ | 2.7723 | Average coffee yield per tree | Mahintana Dataset (2021) |
| $I$ | 1.025 | Average number of trees in 1 hectare | Mahintana Dataset (2021) |
| $\tau$ | 0.80 | Percentage of bearing trees in 1 hectare | Mahintana Dataset (2021) |
| $\varepsilon$ | 0.05 | Damage rate incurred during harvesting | Brando, 2004 (Coffee: Growing, Processing, Sustainable Production ch 2 Harvesting and Green Coffee Processing) |
| $\beta_i$ | [0,1] | Percentage of FC sold to market $i$ | Mahintana Dataset (2021) |
| $\delta_i$ | [0,1] | Percentage of DC sold to market $i$ | Mahintana Dataset (2021) |
| $\mu$ | 0.45 | FC to DC conversion rate | Wintgens, 2004 (Coffee: Growing, Processing, Sustainable Production) |
| $\sigma_i$ | [0,1] | Percentage of GCB sold to market $i$ | Mahintana Dataset (2021) |
| $\theta$ | 0.20 | FC to GCB conversion rate | Philippine Cacao Roadmap |

***Developing the model***

In developing the SCND model for coffee chain, we assumed that the model farmer or the regular farmer is a farmer owning a farm size of one ha. This is to normalize the parameter values used in the model. We first formulated an equation that describes coffee production $C_t$. The total amount of fresh cherries harvested in 1 cycle is given by:

$$C_t = \Lambda_t I \tau (1 - \varepsilon) \qquad (1)$$

where $\Lambda_t$ is the coffee yield per tree multiplied by the average number of coffee trees in 1 hectare ($I$) and the percentage of bearing trees per hectare ($\tau$), this is affected by damage rate $\varepsilon$. According to the literature, farmers using a strip-picking method will start to harvest if 95% of the coffee cherries are ripe, hence, we set the damage rate $\varepsilon = 0.05$.

*Corresponding Author: groguis@up.edu.ph; (082) 293 0303

The harvested coffee cherries of the farmer will be allocated into 3 since coffee farmers in Sultan Kudarat usually sell 3 types of coffee products to markets $i$ = 1,2,..,5 for Nestle, local traders grower association, direct selling, and other markets, respectively.

The allocation of the three coffee products to markets $i = 1, 2, 3, 4, 5$ is given by

$$f_{(i,t)} = \beta_i C_t \tag{2.1}$$

$$R_t = C_t - \sum_{i=0}^{5} f_{(i,t)} \tag{2.2}$$

$$d_{(i,t)} = \delta_i R_t \tag{2.3}$$

$$G_t = R_t - \sum_{i=0}^{5} d_{(i,t)} \tag{2.4}$$

$$g_{(i,t)} = \sigma_i R_t \tag{2.5}$$

where $f_{(i,t)}$ is the allocation of fresh cherries to markets $i$. This is calculated by multiplying the amount of harvested coffee cherries $C_t$ by the percent allocation $\beta_i$ to markets $i = 2, 5$ (for $i = 1, 3, 4$, $\beta_i = 0$). To calculate the coffee cherries to be processed into dried cherries, the remaining cherries $R_t$ are calculated using equation 2.2 and then multiplied by the percent allocation $\delta_i$ to markets $i = 2, 3, 4, 5$ ($\delta_1 = 0$). Lastly, to calculate for the coffee cherries to be processed into green coffee beans, the remaining coffee cherries after allocating to the markets for FC and DC ($G_t$) are multiplied by the percent allocation $\sigma_i$ to markets $i = 1, 2, 3, 4, 5$.

*Corresponding Author: groguis@up.edu.ph; (082) 293 0303

the amount of dried cherries $d^*_{(i,t)}$ and green coffee beans $g^*_{(i,t)}$ after drying and dehulling, respectively is calculated as

$$d^*_{(i,t)} = d_{(i,t)}\mu \qquad\qquad (3.1)$$

$$g^*_{(i,t)} = g_{(i,t)}\theta \qquad\qquad (3.2)$$

where $\mu$ and $\theta$ are the conversion rates from fresh cherry to dried cherry and fresh cherry to GCB, respectively.

The total sales gained by the farmer are calculated using Equation 4.

$$P_{(i,t)} = \omega_i f_{(i,t)} + \pi_i d^*_{(i,t)} + \rho_i g^*_{(i,t)} \qquad\qquad (4)$$

where $\omega_i, \pi_i,$ and $\rho_i$ are the average prices of FC, DC, and GCB per kg sold to market $i = 1, 2, 3, 4, 5$, respectively.

The total profit gained by farmers in selling coffee products is calculated using Equation 5.

$$P(t) = \left(\sum_{i=1}^{5} P_i(t)\right) - \chi_j, \, for \, j = 1, 2, 3 \qquad\qquad (5)$$

where $\chi_j$ is the cost incurred when selling $j = 1, 2, 3$ coffee products, FC, DC, GCB, respectively. The costs incurred in selling fresh cherries are the production cost, harvesting cost, and transportation cost. If selling DC, additional cost for drying is incurred. and when selling GCB, all the above costs are incurred plus the drying cost.

*Corresponding Author: groguis@up.edu.ph; (082) 293 0303

Table 3 shows the parameter values used in running the model, the sources of these values are secondary data obtained from Mahintana Foundation Inc., through a partnership with the Inclusive Value Chain Project in UP Mindanao, and a literature review. Table 4 displays the actual cases happening to the coffee farmers in Sultan Kudarat when marketing their produce to different market outlets that will be used for the scenario analysis.

Table 4. Summary of the cases used in scenario analysis.

| Cases | Type of Coffee sold | Market* | Variable | Allocation of coffee type sold to market i | $\sum\limits_{i=1}^{5} \beta_i$ | $\sum\limits_{i=1}^{5} \delta_i$ | $\sum\limits_{i=1}^{5} \sigma_i$ |
|---|---|---|---|---|---|---|---|
| 1 | FC | 2 | $\beta_2$ | 1 | 1 | 0 | 0 |
| 2 | FC | 5 | $\beta_5$ | 1 | 1 | 0 | 0 |
| 3 | DC | 2 | $\delta_2$ | 1 | 0 | 1 | 0 |
| 4 | DC | 3 | $\delta_3$ | 1 | 0 | 1 | 0 |
| 5 | DC | 4 | $\delta_4$ | 1 | 0 | 1 | 0 |
| 6 | DC | 5 | $\delta_5$ | 1 | 0 | 1 | 0 |
| 7 | GCB | 1 | $\sigma_1$ | 1 | 0 | 0 | 1 |
| 8 | GCB | 2 | $\sigma_2$ | 1 | 0 | 0 | 1 |
| 9 | GCB | 3 | $\sigma_3$ | 1 | 0 | 0 | 1 |
| 10 | GCB | 4 | $\sigma_4$ | 1 | 0 | 0 | 1 |
| 11 | GCB | 5 | $\sigma_5$ | 1 | 0 | 0 | 1 |
| 12 | DC | 2,3,4,5 | $\delta_3\delta_2 + \delta_4 + \delta$ | 70% 30% | 0 | 1 | 0 |
| 13 | GCB | 1,2,3,4 | $\sigma_3\sigma_1 + \sigma_2 + \sigma_3 +$ | 70% 30% | 0 | 0 | 1 |
| 14 | FC | 2,5 | $\beta_2 + \beta_5$ | [0,1] | 1 | 0 | 0 |

*Corresponding Author: groguis@up.edu.ph; (082) 293 0303

* 1 – Nestle, 2 – Local traders, 3 – Grower association, 4 – Direct selling, 5 – Other markets

## RESULTS

Figure 4 illustrates the profit of farmers selling 100% of their fresh cherries (FC), dried cherries (DC), and green coffee beans (GCB) under different scenarios (Cases 1-11), considering the prices of coffee products (Table 2). Assuming that the annual coffee yield per tree and the average prices of coffee products in the different markets are constant, the analysis shows that in Case 1, farmers can gain positive annual profit per hectare by selling all FC to a market with an average selling price of around Php 32.00 per kg. On the other hand, if the price of FC per kg is Php 12.00 (Case 2), farmers will incur a negative profit.

For cases 3-6, model farmers can gain a positive average annual profit per hectare by selling all DC to different market outlets that purchase dried cherries. However, suppose that farmers engage in further processing, such as dehulling to produce and sell green coffee beans, incurring an additional dehulling cost of around Php 2.51 per tree. In that case, they will gain negative profit (Cases 7-11). One possible explanation for this negative result is that the average selling price of GCB shows minimal difference compared to the average selling price of DC, despite having additional costs incurred during the dehulling process of the former.

*Corresponding Author: groguis@up.edu.ph; (082) 293 0303

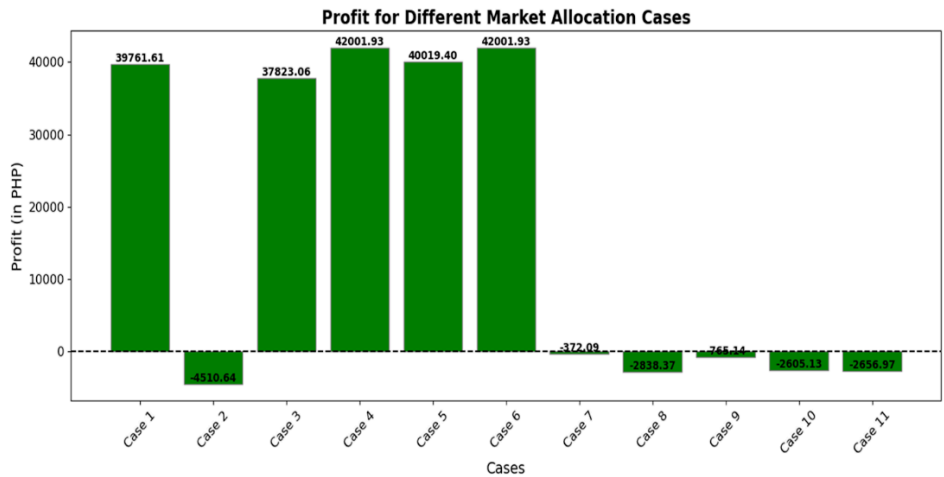

Figure 4. Profit of Model Farmers for Cases 1-11.

For cases 12 and 13, we assumed that members of the growers' association producing DC and GCB, respectively, are obligated to sell 70% of their produce to their association, while they are free to sell the remaining 30% to different markets. The analysis reveals that members of the grower associations selling DC can gain positive annual profit per hectare by allocating the remaining 30% of their DC to Other Markets (Figure 5). However, selling 30% of GCB to any market results in a negative profit.

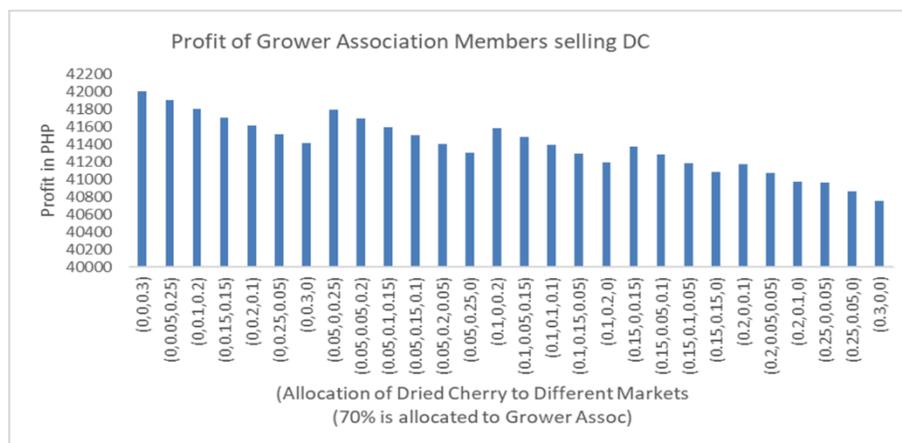

Figure 5. Graph for Case 12

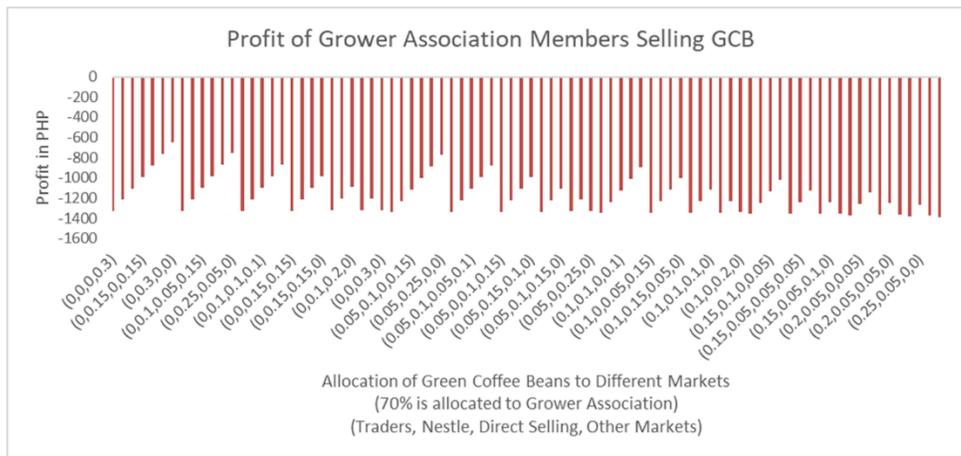

Figure 6. Graph for Case 13.

Additional analysis was performed by exploring various scenarios, such as examining the profit of farmers when selling FC to both local traders and Other Markets. The aim was to determine the percentage allocation per market to achieve a positive profit. Based on the findings, assuming the prices of FC are Php 32.00 and Php 12.00 for local traders and Other Markets, respectively, farmers can gain a minimum positive annual profit of Php 2,130.19 by allocating at least 15% of FC to local traders and the remaining to the Other Market. However, if farmers allocate at least 90% of FC to the Other Market, the profit becomes negative (Php = -83.41).

Additionally, the prices of GCB were simulated to identify the minimum price required for farmers to gain a positive profit (Figure 8). A similar simulation was also performed for dehulling costs, determining the maximum dehulling cost of farmers selling GCB to achieve a positive profit (Figure 9).

The result shows that farmers will gain a positive minimum annual profit of Php 314.67 selling green coffee beans if the price is at least Php 77.00. Furthermore, given that the selling price of

*Corresponding Author: groguis@up.edu.ph; (082) 293 0303

GCB offered by Nestle is Php 75.41, the annual profit of farmers will become positive (Php 46.11) if the dehulling cost is reduced to Php 2.00 per tree.

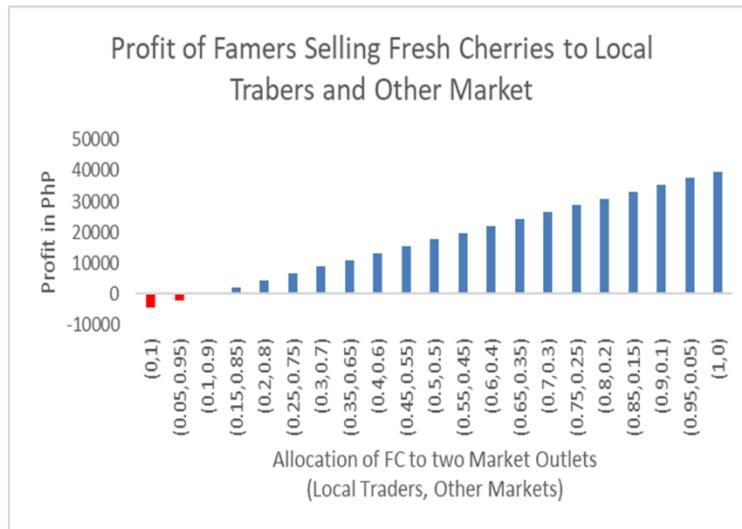

Figure 7. Profit of farmers selling FC to both Local Traders and Other Markets

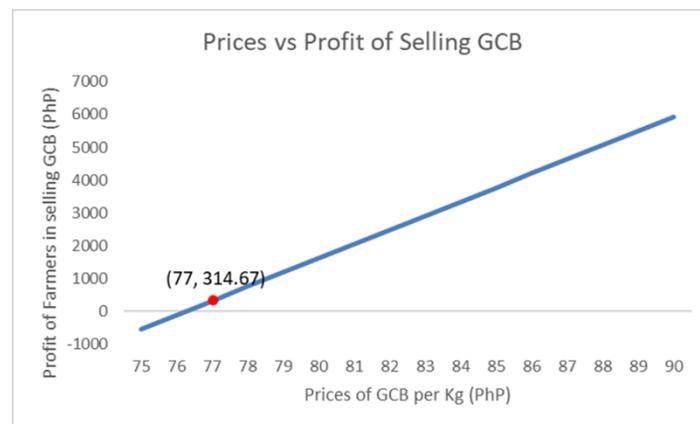

Figure 8. Simulation results to determine the minimum price of GCB to gain positive profit.

*Corresponding Author: groguis@up.edu.ph; (082) 293 0303

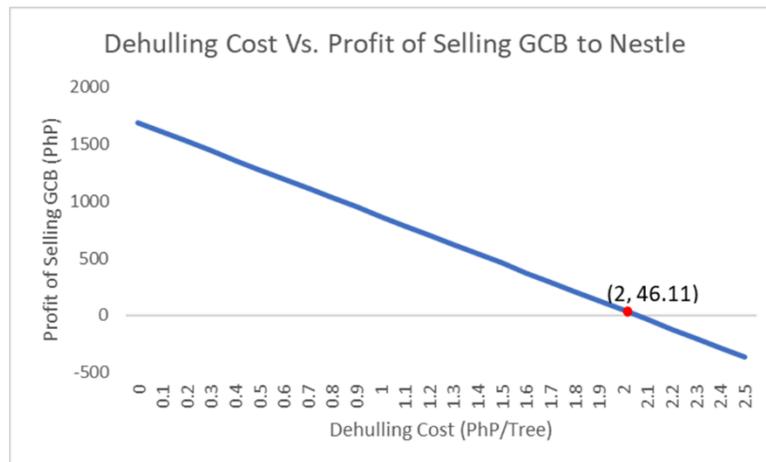

Figure 9. Simulation results to determine the maximum dehulling cost to gain positive profit

DISCUSSION

This study aims to analyze the coffee chain in Sultan Kudarat based on the usual practices of coffee farmers cultivating robusta coffee variety using SCND method. Robusta coffee variety is the most produced coffee bean in the country. This coffee variety is more resistant to diseases, produces better yield than other coffee varieties, consists of more caffeine, and is widely used in espresso blends and instant coffee production (Alves et al, 2009). It was predicted that in 2025, the coffee consumption of Filipinos will grow by an average of 8.1% a year from 2022 to 2025. This indicates that there is an increase in demand for coffee products. Moreover, the study of Consignado and Dimaculangan (2022) also indicates that new coffee enterprises can quickly enter the market. Hence, the entrance/introduction of these new players in the coffee chain is an opportunity for coffee farmers to market their products due to the increase in coffee consumers.

However, the coffee farmers in Sultan Kudarat seem to gain lower profit especially when selling green coffee beans despite having five classified coffee buyers. Therefore, this study particularly aims to identify the optimal market choices of coffee farmers based on their usual practices and resources to help in their decision-making process in marketing their coffee products. By this, scenario analysis was performed to identify which market should the coffee farmers allocate their products to gain the optimal profit.

Given the practices of coffee farmers in Sultan Kudarat, the analysis shows that further processing the coffee cherries to produce green coffee beans is not recommended for farmers since they will incur additional costs but the prices are just the same as the average price in selling dried coffee beans. According to Jun's (2022) dissertation about the value addition in coffee production in Uganda, there is a better benefit in the profitability of coffee farmers if they sell dried cherries which only includes drying fresh cherries usually using the sun-drying method.

 However, the price of coffee beans is directly related to its quality. According to the coffee farmers in Jakarta from the study of FAO (2005), the price incentives given to the farmers by the buyers are not sufficient to compensate for the former's costs and risks incurred when producing better coffee quality. Hence, the reason for the decrease in the quality of coffee produced by farmers (Susila, 2006).

Most farmers in Sultan Kudarat used a natural drying method. This method is relatively cheaper, requires less labor than other methods in coffee processing, and carries a lower risk of quality degradation during processing (Jun 2020). Moreover, value addition to the coffee would imply a higher price. For instance, the price of dried cherries is relatively higher than fresh cherries and the price of green coffee beans is higher than the two coffee products. But, in Sultan Kudarat, this is not the. Based on the secondary data from MFI (2021), there is no big difference between the average price of dried cherries (PhP 73.42) and green coffee beans (Php 71.99). The average price of dried cherries is slightly higher than that of green coffee beans. This may explain why selling dried cherries is a better option for farmers to gain the optimal profit compared to selling green coffee beans.

*Corresponding Author: groguis@up.edu.ph; (082) 293 0303

It was suspected that the dehulling cost for producing green coffee beans is a factor that affects negative profit. Hence, the profit of farmers was analyzed based on the costs incurred in the dehulling process to identify the minimum cost of this process so that the profit of coffee farmers becomes positive.

Moreover, knowing the minimum price of green coffee beans to gain positive profit is an interesting aspect to investigate. Specifically, the case when farmers sell GCB to Nestle Philippines - one of the big buyers of coffee beans in the Philippines, was used to explore new case, since this is the market option that has the closest profit to zero. The analysis shows that when the price of green coffee beans starts to reach at least Php 77.00, it will generate a positive profit of Php 314.67. If the price of green coffee beans in Sultan Kudarat is the same as the global average market price of US$ 1.98 (Php 97.55) (Statista, 2021), then coffee farmers will gain higher positive profits. Furthermore, if the dehulling cost decreases to Php 2.00, farmers' profits will increase significantly.

Policymakers should invest in supporting the coffee farmers in value addition since it is evident that engaging in value-adding in coffee commodities tends to increase the profit of coffee farmers. A study of Jun (2020) shows that factors such as an increase in coffee farm size, and education of coffee farmers have positive impacts on coffee value addition. Farmers are more likely to adopt innovative practices if they have a bigger land area planted with coffee (Feder 1980; Boahene 1999). Bigger coffee production would be advantageous to add value that, in turn, would lead to a higher income for the farmers (Jun 2020). However, the majority of the coffee farmers in Sultan Kudarat are small-scale, hence, policymakers should strongly support the F2C2 program implemented by the Department of Agriculture.

*Corresponding Author: groguis@up.edu.ph; (082) 293 0303

Additionally, educated farmers tend to adopt improved practices for their coffee produce (Musebe, 2007). This is because they are knowledgeable to avoid the deterioration of the quality of coffee and produce higher quality value-added products, which is expected to bring higher prices (Jun 2020). Policymakers can implement various training and seminars for coffee post-production and value-adding to help farmers the proper way of processing coffee cherries to produce high-quality beans.

CONCLUSION

In conclusion, based on the data and assumptions used in the scenario analysis, it is recommended to sell fresh cherries or dried cherries to gain positive profits, considering the farmer's current resources and farming practices. This approach was inspired by the developed SCND model for the Cavendish banana commodity in the Davao Region, which serves as a diagnostic tool for assessing the profitability of a commodity in different market options. Therefore, the developed model can be modified and can be applied to regular coffee farms and other commodities, and the simulation results can offer recommendations to smallholder farmers by identifying the optimal market choices to increase their profitability.